\documentclass[reqno,12pt]{article}
\usepackage{amsmath,amsfonts,amssymb,amstext,amscd,eucal}
\usepackage[all]{xy}
\usepackage{mathtools}
\usepackage{slashed}
\usepackage{float}
\usepackage{simplewick}
\usepackage{xcolor,colortbl}

\newcommand{\mathsym}[1]{{}} 
\usepackage[colorlinks=true,
            linkcolor=blue,
            urlcolor=blue,
            citecolor=blue]{hyperref}
\usepackage{enumerate}

\usepackage{graphicx}
\usepackage{bm}
\usepackage{cite,hyperref}

\DeclareMathAlphabet{\pazocal}{OMS}{zplm}{m}{n}
\usepackage[toc,page]{appendix}
\usepackage[active]{srcltx}
\makeatletter \@addtoreset{equation}{section}

\makeatletter\renewcommand\section{\@startsection {section}{1}{\z@}%
                                   {-3.5ex \@plus -1ex \@minus -.2ex}
                                   {2.3ex \@plus.2ex}%
                                   {\normalfont\large\bfseries}}
\renewcommand\subsection{\@startsection{subsection}{2}{\z@}%
                                     {-3.25ex\@plus -1ex \@minus -.2ex}%
                                     {1.5ex \@plus .2ex}%
                                     {\normalfont\bfseries}}

\newcommand{\be}{\begin{equation}}
\newcommand{\ee}{\end{equation}}
\newcommand{\bea}{\begin{eqnarray}}
\newcommand{\eea}{\end{eqnarray}}
\newcommand{\bse}{\begin{subequations}}
\newcommand{\ese}{\end{subequations}}
\newcommand{\beqa}{\begin{eqnarray}}
\newcommand{\eeqa}{\end{eqnarray}}
\newcommand{\beqar}{\begin{eqnarray*}}
\newcommand{\eeqar}{\end{eqnarray*}}
\newcommand{\bi}{\begin{itemize}}
\newcommand{\ei}{\end{itemize}}
\newcommand{\bn}{\begin{enumerate}}
\newcommand{\en}{\end{enumerate}}

\newcommand{\ba}{\begin{array}}
\newcommand{\ea}{\end{array}}
\newcommand{\bc}{\begin{center}}
\newcommand{\ec}{\end{center}}

\definecolor{darkgreen}{rgb}{0,0.3,0}
\definecolor{darkblue}{rgb}{0,0,0.3}
\definecolor{darkred}{rgb}{0.7,0,0}
\definecolor{VioletRed4}{rgb}{0.55,0.13,0.32}
\definecolor{VioletRed}{rgb}{0.82,0.13,0.56}
\definecolor{VioletRed2}{rgb}{0.93,0.23,0.55}

\begin{document}
\setcounter{footnote}{0}
\renewcommand{\baselinestretch}{1.05}
\newcommand{\email}[1]{\footnote{\href{mailto:#1}{#1}}}

\title{\bf\Large{Induced CP-violation in the Euler-Heisenberg Lagrangian}}
\author{\bf{M.~Ghasemkhani}\email{ghasemkhani@ipm.ir} $^{a}$, \bf{V.~Rahmanpour}\email{v.rahmanpour@mail.sbu.ac.ir} $^{a}$, R.~Bufalo\email{rodrigo.bufalo@ufla.br} $^{b}$,\\
\bf{M.N. Mnatsakanova\email{mnatsak@theory.sinp.msu.ru} $^{c}$ and A.~Soto\email{arsoto1@uc.cl} $^{d}$}\\\\
\textit{\small$^a$ Department of Physics, Shahid Beheshti University, 1983969411, Tehran, Iran}\\
\textit{\small$^b$ Departamento de F\'isica, Universidade Federal de Lavras,}\\
\textit{ \small Caixa Postal 3037, 37200-900 Lavras, MG, Brazil}\\
  \textit{ \small$^c$ Skobeltsyn Institute of Nuclear Physics, Lomonosov Moscow State University,}\\
 \textit{ \small Moscow, Russia}\\
\textit{\small$^d$ School of Mathematics, Statistics and Physics, Newcastle University,}\\
 \textit{ \small Newcastle upon Tyne, NE1 7RU, UK}\\
}
\maketitle
\begin{abstract}
In this paper, we examine the behaviour of the Euler-Heisenberg effective action in the presence of a novel axial coupling among the gauge field and the fermionic matter.
This axial coupling is responsible to induce a CP-violating term in the extended form of the Euler-Heisenberg effective action, which is generated naturally through the analysis of the box diagram.

\noindent
However,
this anomalous model is not a viable extension of QED, and we explicitly show that the induced CP-violating
term in the Euler-Heisenberg effective Lagrangian is obtained only by adding an axial coupling to the ordinary QED Lagrangian.
In order to perform our analysis, we use a parametrization of the vector and axial coupling constants, $g_{v}$ and $g_{a}$, in terms of a new coupling $\beta$.
Interestingly, this parametrization allows us to explore a hidden symmetry under the change of $g_{v}\leftrightarrow g_{a}$ in some diagrams.
This symmetry is explicitly observed in the analysis of the box diagram, where we determine the $\lambda_i$ coefficients of $\pazocal{L}_{\rm ext.}^{\rm \small EH}=\lambda_{1}\pazocal{F}^{2}+\lambda_{2}\pazocal{G}^{2}+\lambda_{3}\pazocal{F}\pazocal{G}$, specially the coefficient $\lambda_3$ related with the CP-violating term due to the axial coupling.
As a phenomenological application of the results, we
compute the relevant cross section for the light by light scattering through the extended Euler-Heisenberg effective action.
\end{abstract}
\newpage
\section{Introduction}

Discrete symmetries play a major role in particle physics, $C$ as charge conjugation, $P$ as parity transformation and $T$ as time reversal.
They eventually led to Pauli's CPT theorem, a cornerstone of particle physics \cite{Sozzi:2008zza}.
Naturally, there are many important and interesting examples where these discrete symmetries (or a combination of them ) are violated, e.g. CP violation in the electroweak theory and the strong CP problem in QCD \cite{Sozzi:2008zza}, P violating effects in planar QFT \cite{Deser:1981wh}.

As a direct consequence of these fundamental results, there is currently interest in examining the validity of the CPT theorem \cite{Kostelecky:2008ts}, as well as other combinations of these discrete symmetries in phenomenological analysis.
As a part of this endeavour, a scenario where these violation effects could be analyzed is within the
nonlinear electrodynamics, in which implications of these nonlinear electromagnetic effects are being examined by several experiments \cite{Cadene:2013bva,DellaValle:2014xoa,ATLAS:2017fur,ATLAS:2019azn}.

It is well known that the parity breaking is a rich scenario to introduce some important phenomena, even inducing massive modes, regarding the behaviour of the electromagnetic field in three dimensions \cite{Deser:1981wh}.
For instance, one can cite to the quantum Hall effect as an important physical application where the topological effects of electromagnetism are the framework to describe this physical phenomena \cite{ezawa}.
Hence, one should expect that the violation of parity would be also interesting in the nonlinear corrections to the four-dimensional electrodynamics.

Nonlinear electrodynamics has a long history, from the early proposals of the light-by-light scattering in QED  \cite{Halpern,heisenberg,euler-kockel,heisenberg-euler,karplus-neuman,Fedotov:2022ely} to the conceptual proposal of light-by-light scattering in ultraperipheral heavy-ion collisions  \cite{dEnterria:2013zqi} as well as its experimental verification by ATLAS Collaboration \cite{ATLAS:2017fur,ATLAS:2019azn}.
These nonlinear terms in the electromagnetic field equations can be understood in terms of quantum effects in the context of  effective field theories \cite{Dittrich:2000zu}, resulting in the phenomena of quantum self-coupling of electromagnetic waves in the vacuum.

Interestingly, the experimental verification of the light-by-light scattering is important since it is now serving as a laboratory to investigate physics beyond the standard model scenarios. These models predict new particles that couple predominantly to photons, for instance, the search for the axion-like particles \cite{ATLAS:2020hii,Goncalves:2021pdc}.
On the theoretical side, we have seen in recent years great interest in the study of further nonlinear phenomena involving the physics beyond the standard model of gauge bosons, exploring new possibilities and the novel (quantum) behaviour of the electromagnetic waves \cite{Ayon-Beato:1998hmi,Bonora:2016otz,Bonora:2017ykb,Preucil:2017,Quevillon:2018mfl,Yamashita:2017,Fan:2017sxk,Karbstein-2021,Horvat:2020ycy,Gorghetto:2021luj,Rizzo:2022qan}.
Our proposal lies within this context, where we wish to consider some generalization of these previous studies and consider the implications of CP-violating effects into the photon dynamics.


As we know, the CP-violating effects in the four photon interactions do not exist in the
Standard Model at tree-level.
However, there are some sources of photon interactions via
CP-violation in terms of multi-loop level, arising from the weak interactions (CP-violating phase of the CKM
matrix), which are negligibly small \cite{Gorghetto:2021luj,Millo:2008ug}.

Therefore, for the perturbative generation of the CP-violating term in the photon sector, it is
necessary that, at least, one of these discrete symmetries (C or P) is broken.
To this end, we can add new couplings related with physics beyond the standard model.
These couplings are necessarily axial and therefore both C and P symmetries in the photon-matter couplings are automatically broken.
Indeed, only fermionic bilinear covariants such as axial-vector, axial-tensor, etc, coupling with the photon can produce the CP-violating term in the photon sector, see \cite{Gorghetto:2021luj,Ghasemkhani:2022mqw} for further details.
This means that only anomalous fermionic models are related with the CP-violating part of the photon sector.
As discussed in \cite{Ghasemkhani:2022mqw}, the CP-violating term in the photon
sector is only obtained from physically nonviable anomalous models, and not from any fundamental field theory.
As a continuation of the discussion presented in \cite{Ghasemkhani:2022mqw}, we intend to explicitly show how these anomalous couplings induce a CP-violating term in the Euler-Heisenberg effective action.


We start Sec.~\ref{sec2} by establishing the main aspects of the considered model, with a gauge field that possesses both vector and axial couplings, and its related symmetries.
Furthermore, we present some definitions regarding the effective action formalism.
In Sec.~\ref{sec3}, we discuss some details involving the (low-energy) Euler-Heisenberg (EH) Lagrangian.
In particular, we focus in developing the four-photon scattering matrix, and how to consider the different parity preserving and parity violating contributions to the amplitude.
Our main analysis is presented in Sec.~\ref{sec5}, where the evaluation of the lowest-order contribution to the box diagram (with four photon legs) is fully considered.
It is important to emphasize that, since this amplitude needs to be regularized, we follow the 't Hooft-Veltman rule to perform algebraic manipulations with $\gamma_5$ within the dimensional regularization method.
Moreover, in Sec.~\ref{sec6} we analyze our results related with the $\lambda_i$ coefficients of the (parity-violating) Euler-Heisenberg effective action and discuss some particular issues.
In Sec.~\ref{sec7}, as a phenomenological application of our results, we compute the relevant cross section for the light by light scattering through the extended Euler-Heisenberg effective action. Finally, we present our conclusions and final remarks in Sec.~\ref{conc}.

\section{The model and main features}
\label{sec2}

In this section, we introduce the model and fix our notation.
The parity-violating extension of the QED minimal coupling for the Dirac fermions in the presence of an external gauge field is given by the following action
\begin{equation}
{\pazocal{S}}_{\psi}=\int d^{4}x~\bar{\psi}(x)\Big[\gamma^{\mu}\big(i\partial_{\mu}-(g_{v}+g_{a}\gamma_{5})A_{\mu}\big)-m\Big]\psi(x),
\label{eq:a1}
\end{equation}
where the interacting Lagrangian density is
\begin{equation}
{\pazocal{L}}_{int}=-\bar\psi\gamma^{\mu}(g_{v}+g_{a}\gamma^{5})A_{\mu}\psi,
\label{eq:a2}
\end{equation}
where $g_{v}$ and $g_{a}$ refer to the coupling of the external gauge field to the vector and axial vector current, respectively.
We observe that the axial part of the Lagrangian \eqref{eq:a2} violates the parity (P) and charge conjugation (C) symmetry, which is odd under P and C.
Thus, unlike the usual QED, this model does not respect the parity and charge conjugation symmetry, but also CP conjugation is satisfied.
An obvious consequence of the C-violation is that the Furry theorem is no more satisfied in this model.
Another important comment is in regard of the anomalous nature of this coupling. Actually, this axial-vector coupling originates the very same Adler-Bardeen anomaly known from the usual vector coupling of the photon \cite{Balachandran:1981cs,Einhorn:1983nv}.

In order to perform our perturbative analysis, we introduce a parametrization to consider the parity-conserving (vector) and the parity-violating (axial-vector) coupling constants in a simpler fashion
\begin{equation}
g_{v}+g_{a}\gamma^{5}=\beta e^{\alpha\gamma^{5}},
\label{eq:a2-2}
\end{equation}
so that it is readily obtained
\begin{equation}
g_{v}=\beta\cosh\alpha,\quad g_{a}=\beta\sinh\alpha,\quad \beta^{2}=g_{v}^{2}-g_{a}^{2}.
\label{eq:a3}
\end{equation}
It is important to remark that in this new notation, the presence of a new symmetry  $g_{v}\leftrightarrow g_{a}$ is verified in the $n$-point function of the gauge field, see comment below.

In terms of this parametrization, the Lagrangian density \eqref{eq:a1} is rewritten as
\begin{equation}
{\pazocal{L}}_{\psi}=\bar{\psi}(x)\Big[\gamma^{\mu}\big(i\partial_{\mu}-\beta e^{\alpha\gamma^{5}}A_{\mu}\big)-m\Big]\psi(x).
\label{eq:a4}
\end{equation}
As a result, the Feynman rule for the fermion-photon interaction is simply given by $-i\beta\gamma^{\mu}e^{\alpha\gamma^{5}}$, having the same structure as the usual QED with the additional factor $e^{\alpha\gamma^{5}}$.

In regard to the symmetries of the Lagrangian \eqref{eq:a4}, we observe that, with massless fermions, it is invariant under the following gauge transformation
\begin{equation}
\psi(x)\rightarrow e^{i\beta e^{\alpha\gamma^{5}}\theta(x)}\psi(x),\quad
\bar\psi(x)\rightarrow\bar\psi(x)~e^{i\beta e^{\alpha\gamma^{5}}\theta(x)},\quad A_{\mu}(x)\rightarrow A_{\mu}(x)-\partial_{\mu}\theta(x).
\label{eq:a5}
\end{equation}
Moreover, we notice that in the case of $g_{a}\to 0$ and $g_{v}\to 0$, the usual vector gauge transformation, $U_{V}(1)$, and the axial gauge transformation, $U_{A}(1)$, are restored, respectively\footnote{As a matter of fact, we also observe the presence of this axial vector current for the photon in a non-hermitian and PT symmetric extension of QED \cite{Alexandre:2015kra}, where this coupling is related with the (pseudoscalar) mass term $\bar{\psi} \gamma_5 \psi $, and it is necessary in order to preserve the gauge symmetry.}

The one-loop effective action $\Gamma$ for the gauge field $A_{\mu}$ is defined as follows
\begin{equation}
\Gamma[A]=-i\textrm{Tr}\ln\Big(i\slashed{\partial}-\beta e^{\alpha\gamma^{5}}\slashed{A}-m\Big),
\label{eq:a7}
\end{equation}
or equivalently in a convenient form for the perturbative analysis
\begin{equation}
\Gamma[A]=\sum_{n=1}^{\infty}\int d^{4}x_{1}\ldots \int d^{4}x_{n}~\Gamma^{\mu_{1}\ldots \mu_{n}}
(x_{1},\ldots, x_{n})~A_{\mu_{1}}(x_{1})\ldots A_{\mu_{n}}(x_{n}).
\label{eq:a8}
\end{equation}
Here, $\Gamma^{\mu_{1}\ldots \mu_{n}}$ is the $n$-point function of the gauge field which is defined as
\begin{equation}
\Gamma^{\mu_{1}\ldots \mu_{n}}(x_{1},\ldots, x_{n})=-\frac{\beta^{n}}{n}\int\prod_{i=1}^{n}\frac{d^{4}p_{i}}{(2\pi)^{4}}
~\delta\left(\sum\limits_{i=1}^{n} p_{i}\right)~e^{i\sum\limits_{i=1}^{n}p_{i}.x_{i}}~
\Xi^{\mu_{1}\ldots \mu_{n}}\left(p_{1},\ldots,p_{n}\right),
\label{eq:a9}
\end{equation}
where the overall minus sign comes from the fermionic loop and $\Xi^{\mu_{1}\ldots \mu_{n}}$ indicates the amplitude of a graph with $n$ external photon legs.

As a final remark, we can realize from eq.~\eqref{eq:a9} that $\Gamma^{\mu_{1}\ldots \mu_{n}}$ is proportional to the following power of the new (parametrized) coupling constant $\beta^{n} = (g_{v}^{2}-g_{a}^{2})^{\frac{n}{2}} $.
However, the dependence of $\Gamma^{\mu_{1}\ldots \mu_{n}}$ on the coupling constants $g_{v}$ and $g_{a}$ does not arise only from $\beta^{n}$ since we have a factor $e^{\alpha\gamma^{5}}$ for each vertex (included in $\Xi^{\mu_{1}\ldots \mu_{n}}$) that depends on $g_{v}$ and $g_{a}$. As we shall see in Sec.~\ref{sec5}, in the case of interest $n=4$, we have terms proportional to $\beta^{4}e^{4\alpha\gamma^{5}}$ and $\beta^{4}e^{2\alpha\gamma^{5}}$ within eq.~\eqref{eq:a9} as
 \begin{align}
 \beta^{4}e^{4\alpha\gamma^{5}}&=(g_{v}^{4}+g_{a}^{4}+6g_{v}^{2}g_{a}^{2})+4(g_{v}^{3}g_{a}+g_{v}g_{a}^{3})\gamma^{5},\\
 \beta^{4}e^{2\alpha\gamma^{5}}&=(g_{v}^{4}-g_{a}^{4})+2(g_{v}^{3}g_{a}-g_{v}g_{a}^{3})\gamma^{5}.
 \end{align}
 We can straightforwardly notice that $ \beta^{4}e^{4\alpha\gamma^{5}}$ and $ \beta^{4}e^{2\alpha\gamma^{5}}$ are completely symmetric and anti-symmetric under the exchange of $g_{v} \leftrightarrow g_{a}$, respectively.
 Nevertheless, only terms proportional to $ \beta^{4}e^{4\alpha\gamma^{5}}$ will contribute to the extended Euler-Heisenberg Lagrangian as shown in Sec.~\ref{sec5}; although finite, the remaining parts contribute to different effective action other than the Euler-Heisenberg one.
 Therefore, this result allows us to undercover that  the effective Lagrangian is symmetric under $g_{v} \leftrightarrow g_{a}$.

\section{Construction of the extended Euler-Heisenberg Lagrangian}
\label{sec3}

In order to construct the one-loop effective action for the description of the light by light scattering, some comments are in order.
There are strong constraints such as gauge and Lorentz invariance that should be considered in this construction.
Accordingly, a gauge invariant expression to describe the four-photon interaction can be built from the field strength tensor $F_{\mu\nu}$ with proper contractions, leading to a Lorentz scalar Lagrangian.
Moreover, since our interaction term in \eqref{eq:a2} violates parity, we shall have an additional term in the extended EH Lagrangian, which is absent in its usual version.

Based on these comments, there are three kinds of terms with mass dimension 8 that respect both gauge and Lorentz invariance, as well as the parity violation, as below
\begin{align}
\pazocal{F}^{2}, \quad
\pazocal{G}^{2},\quad
 \pazocal{F}\pazocal{G},
 \label{eq:b10}
\end{align}
with the definitions
\begin{align}
\pazocal{F}&= F_{\mu\nu}F^{\mu\nu}=-2(\mathbf{E^2-B^2}),\nonumber\\
\pazocal{G}&=  G_{\mu\nu}F^{\mu\nu}=4(\mathbf{E.B}),
\label{eq:b11}
\end{align}
and $G_{\mu\nu}=\frac{1}{2} \varepsilon_{\mu\nu\rho\sigma}F^{\rho\sigma}$ is the dual of the field strength tensor $F_{\mu\nu}=\partial_{\mu}A_{\nu}-\partial_{\nu}A_{\mu}$.
Taking into account these considerations, the full structure of the extended EH Lagrangian can be expressed as the following
\begin{align}
\pazocal{L}_{\rm ext}^{^{\rm EH}}=\lambda_{1}\pazocal{F}^{2}+\lambda_{2}\pazocal{G}^{2}+\lambda_{3}\pazocal{F}\pazocal{G},
\label{eq:b12}
\end{align}
where $\lambda_{1}$, $\lambda_{2}$ and $\lambda_{3}$ are the effective coupling constants with mass dimension $-4$ in a four dimensional space-time.
Under parity, $\mathbf{E} \to - \mathbf{E}$ and $\mathbf{B} \to \mathbf{B}$, the quantities $\pazocal{F}$ and $\pazocal{G}$ transform as scalar and pseudo-scalar, respectively.
Consequently, $\pazocal{F}^{2}$ and $\pazocal{G}^{2}$ are parity-even, whereas $\pazocal{F}\pazocal{G}$ is parity-odd.
This parity-odd term is not present in the ordinary EH action.
Furthermore, since under charge conjugation the field strength changes as $F_{\mu\nu} \to -F_{\mu\nu}$, the full Lagrangian \eqref{eq:b12} is charge-conjugation invariant. Therefore, the quantities $\pazocal{F}^{2}$ and $\pazocal{G}^{2}$ are CP-even (CP-conserving) while $\pazocal{F}\pazocal{G}$ is CP-odd (CP-violating) and hence the full Lagrangian \eqref{eq:b12} is not CP-invariant.

We mention that the Standard Model contains  sources of photon interactions via CP violation at the higher  multi-loop level from the weak interactions (from the phase in the CKM matrix), or from the QCD $\theta$-term, and in both cases they are negligibly small \cite{Millo:2008ug,Gorghetto:2021luj}.
Given the suppression of the standard model contribution, we shall analyze the effects of this CP-odd new coupling, $\lambda_{3}$, in the photon-photon scattering in section \ref{sec7}.

The values of the coupling constants $\lambda_i$ are determined from the one-loop quantum corrections, including the effects of the generalized interaction \eqref{eq:a2}.
To achieve this goal, we determine the total amplitude of the four-photon scattering as a function of the three coupling constants $\lambda_{1}$, $\lambda_{2}$ and $\lambda_{3}$:
first, we will establish the tensor structure of the total amplitude of the process $\gamma\gamma \to \gamma\gamma$ through the Lagrangian \eqref{eq:b12};
second, we will evaluate the  low-energy limit ($p^{2}\ll m^{2}$) of the Feynman amplitudes related with the process $\gamma\gamma \to \gamma\gamma$ by considering the interaction term \eqref{eq:a2}.
At last, we will compare and match the obtained results in both methods, allowing us to fix the values of the coupling constants $\lambda_i$.



Let us decompose the full effective Lagrangian \eqref{eq:b12} into three pieces $\pazocal{L}_{\rm ext}^{^{\rm EH}}= \sum \limits_{i=1} ^3 \pazocal{L}_{i} $, where we have defined
\begin{align}
\pazocal{L}_{1}=\lambda_{1}\pazocal{F}^{2},\quad \pazocal{L}_{2}=\lambda_{2}\pazocal{G}^{2},\quad
\pazocal{L}_{3}=\lambda_{3} \pazocal{F}\pazocal{G}.
\label{eq:b13}
\end{align}
The contribution of the parity-conserving (P.C.) parts, $\pazocal{L}_{1}=\lambda_{1}\pazocal{F}^{2}$ and $\pazocal{L}_{2}=\lambda_{2}\pazocal{G}^{2}$, into the amplitude for four-photon scattering is given by \cite{Preucil:2017}
\begin{equation}
\pazocal{M}_{1}+\pazocal{M}_{2}=\big(\Xi_{(1)}^{\mu\nu\rho\sigma}+\Xi_{(2)}^{\mu\nu\rho\sigma}\big) \left(p_1,p_2,p_3,p_4\right) \varepsilon_{1\mu}\varepsilon_{3\nu}\varepsilon_{4\rho}\varepsilon_{2\sigma}.
\label{eq:b14}
\end{equation}
where $\varepsilon_{i}\equiv \varepsilon (p_i)$ are the polarization vectors corresponding to the $p_i$ photon four-momenta.
In addition, in order to obtain the correct amplitude,
we shall sum over all simultaneous permutations of $(p_1,p_2,p_3,p_4)$ and $(\mu,\sigma,\nu,\rho)$.
As a matter of fact, this corresponds to 24 permutations of the external photon legs which are included in the amplitudes as
\begin{align}
\Xi_{(1)}^{\mu\nu\rho\sigma}=\sum\limits_{i=1}^{24}\Xi_{(1,i)}^{\mu\nu\rho\sigma}(p_1,p_2,p_3,p_4),\quad
\Xi_{(2)}^{\mu\nu\rho\sigma}=\sum\limits_{i=1}^{24}\Xi_{(2,i)}^{\mu\nu\rho\sigma}(p_1,p_2,p_3,p_4),
\label{eq:b15}
\end{align}
where the first contributions are straightforwardly obtained \cite{Preucil:2017}
\begin{align}
\Xi_{(1,1)}^{\mu\nu\rho\sigma}&=4\lambda_{1}\Big[
-(p_1.p_3)(p_2.p_4)g_{\mu\nu}g_{\rho\sigma}+2(p_1.p_3)g_{\mu\nu}p_{4\sigma}p_{2\rho}
 -p_{1\nu} p_{3\mu} p_{4\sigma} p_{2\rho}\Big],
\label{eq:b16} \\
\Xi_{(2,1)}^{\mu\nu\rho\sigma}&=8\lambda_{2}\Big[
(p_1.p_3)(p_4.p_2)g_{\mu\nu}g_{\rho\sigma}
-2(p_1.p_3)g_{\mu\nu}p_{4\sigma}p_{2\rho}
+p_{1\nu}p_{3\mu}p_{4\sigma}p_{2\rho}
+(p_1.p_3)g_{\nu\rho}p_{4\sigma}p_{2\mu}
\cr
&+(p_4.p_2) p_{1\nu}p_{3\rho}g_{\mu\sigma}
+(p_3.p_4)p_{1\nu}p_{2\mu}g_{\rho\sigma}
+(p_1.p_3)p_{4\nu}p_{2\rho} g_{\mu\sigma}
-p_{1\nu}p_{3\rho}p_{4\sigma}p_{2\mu}
\cr
&-(p_1.p_3)(p_4.p_2)g_{\mu\sigma}g_{\nu\rho}-(p_1.p_3)p_{4\nu} p_{2\mu} g_{\rho\sigma}
-(p_3.p_4)p_{1\nu} p_{2\rho} g_{\mu\sigma}\Big].
\end{align}
 After performing the complete permutations in \eqref{eq:b15} and simplifying it, we find that the complete parity-conserving amplitude reads
\begin{align}
\Big(\Xi^{\mu\nu\rho\sigma}\Big)_{P.C}&= 32 (\lambda_{1}-\lambda_{2})\Big[p_{1}^{\sigma}p_{3}^{\mu}p_{3}^{\rho}p_{4}^{\nu}
+p_1^{\sigma}p_3^{\rho}p_4^{\mu}p_4^{\nu}
-p_1^{\nu}p_1^{\rho}p_3^{\sigma}p_4^{\mu}
+p_1^{\rho}p_3^{\sigma}p_4^{\mu}p_4^{\nu}
-p_1^{\nu}p_1^{\rho}p_3^{\mu}p_4^{\sigma}
+p_1^{\nu}p_3^{\mu}p_3^{\rho}p_4^{\sigma}\Big]
\cr
&- 32\lambda_2 \Big[p_1^{\rho}p_1^{\sigma}p_3^{\mu}p_4^{\nu}
+ p_1^{\nu}p_1^{\sigma}p_3^{\rho}p_4^{\mu}
- p_1^{\rho}p_3^{\mu}p_3^{\sigma}p_4^{\nu}
- p_1^{\nu}p_3^{\rho}p_3^{\sigma}p_4^{\mu}
- p_1^{\rho}p_3^{\mu}p_4^{\nu}p_4^{\sigma}
- p_1^{\nu}p_3^{\rho}p_4^{\mu}p_4^{\sigma}\Big]
\label{eq:b17}
\end{align}
 where $\Big(\Xi^{\mu\nu\rho\sigma}\Big)_{P.C}\equiv\Xi_{(1)}^{\mu\nu\rho\sigma}+\Xi_{(2)}^{\mu\nu\rho\sigma}$.
 It is worth mentioning that in eq.~\eqref{eq:b17} only terms without the metric are present, since the rest of the terms cancelled by applying the energy-momentum conservation.

Similarly, we can express the contribution of the parity-violating (P.V) term, $\pazocal{L}_{3}=\lambda_{3}\pazocal{F}\pazocal{G}$, for the amplitude as the following
\begin{equation}
\pazocal{M}_{3}=\Xi_{(3)}^{\mu\nu\rho\sigma} (p_1,p_2,p_3,p_4) \varepsilon_{1\mu}\varepsilon_{3\nu}\varepsilon_{4\rho}\varepsilon_{2\sigma},
\label{eq:b18}
\end{equation}
with the definition
\begin{align}
\Big(\Xi^{\mu\nu\rho\sigma}\Big)_{P.V} = \Xi_{(3)}^{\mu\nu\rho\sigma}=\sum\limits_{i=1}^{24}\Xi_{(3,i)}^{\mu\nu\rho\sigma}(p_1,p_2,p_3,p_4),
\label{eq:b19}
\end{align}
accounting for all possible permutation of the external photon legs.
The first contribution is found as
\begin{align}
\Xi_{(3,1)}^{\mu\nu\rho\sigma}=
8\lambda_{3}\Big(p_{1}^{\nu } p_{3}^{\mu }- g^{\mu\nu}(p_{1}.p_{3})\Big)\epsilon ^{p_{2}p_{4}\rho\sigma },
\label{eq:b20}
\end{align}
where we have defined the condensed notation $\epsilon^{p_{i}p_{j}\rho\sigma}\equiv p_{i\xi}p_{j\eta}\epsilon^{\xi\eta\rho\sigma}$.

Inserting \eqref{eq:b20} back into \eqref{eq:b19} and summing over all remaining permutations, we get
\begin{align}
\Big(\Xi^{\mu\nu\rho\sigma}\Big)_{P.V}&=
32 \lambda_{3}\Big[ p_{1}^{\sigma }  p_{3}^{\mu } \epsilon ^{\nu   p_{3}  p_{4} \rho }+  p_{1}^{\sigma }  p_{4}^{\mu } \epsilon ^{\nu   p_{3}  p_{4} \rho }-  p_{1}^{\rho }  p_{4}^{\sigma } \epsilon ^{\mu  \nu   p_{1}  p_{3}}+  p_{1}^{\rho }  p_{4}^{\mu } \epsilon ^{\nu   p_{1}  p_{3} \sigma } \cr
&+  p_{1}^{\rho }  p_{4}^{\mu } \epsilon ^{\nu   p_{3}  p_{4} \sigma }-  p_{1}^{\nu }  p_{3}^{\sigma } \epsilon ^{\mu   p_{1}  p_{4} \rho }+  p_{1}^{\nu }  p_{3}^{\mu } \epsilon ^{ p_{1}  p_{4} \rho  \sigma }
-  p_{1}^{\nu }  p_{3}^{\mu } \epsilon ^{ p_{3}  p_{4} \rho  \sigma } \cr
& +  p_{3}^{\sigma }  p_{4}^{\nu } \epsilon ^{\mu   p_{1}  p_{4} \rho }+ p_{3}^{\rho }  p_{4}^{\sigma } \epsilon ^{\mu  \nu   p_{1}  p_{3}}+  p_{3}^{\rho }  p_{4}^{\nu } \epsilon ^{\mu   p_{1}  p_{3} \sigma }+  p_{3}^{\rho }  p_{4}^{\nu } \epsilon ^{\mu   p_{1}  p_{4} \sigma }\Big],
\label{eq:b21}
\end{align}
Therefore, the total amplitude for the light by light scattering can be cast as
\begin{align}
\pazocal{M}_{total}=\Big(\Xi^{\mu\nu\rho\sigma}\Big)_{P.C}~\varepsilon_{1\mu}\varepsilon_{3\nu}\varepsilon_{4\rho}\varepsilon_{2\sigma}
+\Big(\Xi^{\mu\nu\rho\sigma}\Big)_{P.V}
~\varepsilon_{1\mu}\varepsilon_{3\nu}\varepsilon_{4\rho}\varepsilon_{2\sigma}.
\label{eq:b22}
\end{align}
We can see in eq.~\eqref{eq:b22} that the first term corresponds to the parity-preserving piece of the QED, while the second piece is the novel part corresponding to the parity-violating effects.
In the next section, we will compute explicitly the total amplitude of the photon-photon scattering by considering the generalized interacting Lagrangian \eqref{eq:a2}.

\section{Perturbative analysis}
\label{sec5}

\begin{figure}[t]
\vspace{-1.2cm}
\includegraphics[height=7.8\baselineskip]{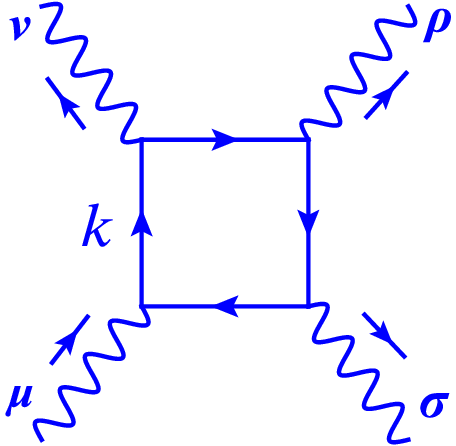}
 \centering\caption{The box diagram contributing to $\langle AAAA\rangle$.}
\label{fig:box}
\end{figure}

In this section, we study the lowest order contribution to the four-photon interaction by performing the one-loop analysis through the underlying theory at the low-energy limit ($p^{2}\ll m^{2}$).
First, we consider the $n=4$ term of the perturbative series \eqref{eq:a8}, corresponding to the so-called box diagram, depicted in Fig.~\ref{fig:box}.
Using the aforementioned Feynman rules, the amplitude of this diagram is given by
\begin{equation}
i\Lambda^{\mu\nu\rho\sigma}_{ (1)}=-\beta^4\int \frac{d^4k}{(2\pi)^4}~
\textrm{tr}\Big[\frac{(\slashed{k}_{1}+m)\gamma^{\mu}e^{\alpha\gamma^{5}}(\slashed{k}+m)\gamma^{\nu}e^{\alpha\gamma^{5}}
(\slashed{k}_{3}+m)\gamma^{\rho}e^{\alpha\gamma^{5}}(\slashed{k}_{34}+m) \gamma^{\sigma}e^{\alpha\gamma^{5}}}
{(k_{1}^2-m^2)(k^2-m^2)(k_{3}^2-m^2)(k_{34}^2-m^2)}
\Big],
\label{eq:a10}
\end{equation}
where $(p_{1}^{\mu},p_{2}^{\sigma},p_{3}^{\nu},p_{4}^{\rho})$ are the momenta of the external legs, satisfying the energy-momentum conservation $p_{1}=p_{2}+p_{3}+p_{4}$. Moreover, we have introduced the condensed notation $k_{i}\equiv k-p_{i}$ and $k_{ij}\equiv k-p_{i}-p_{j}$.

As we have discussed above, in order to find the total amplitude for the 4-point function, we have to consider all of 24 permutations namely,
\begin{equation}
\Lambda^{\mu\nu\rho\sigma}_{\tiny\mbox{total}}=\sum\limits_{i=1}^{24}\Lambda^{\mu\nu\rho\sigma}_{(i)}.
\label{eq:a11}
\end{equation}
To compute the expression \eqref{eq:a10}, first we apply the Feynman parametrization which yields
\begin{equation}
i\Lambda^{\mu\nu\rho\sigma}_{_{(1)}}=-\beta^4\Gamma(4)\int dX \int \frac{d^{\omega}\ell}{(2\pi)^{\omega}}
\frac{{\pazocal{N}}^{\mu\nu\rho\sigma}}
{(\ell^2-\Delta)^4},
\label{eq:a12}
\end{equation}
here we have used $\ell=k-u$ with $u=x p_{1} +(y+z) p_{3} + y p_{4}$ and $\int dX \equiv \int_{0}^{1} dx \int_{0}^{1-x} dy \int_{0}^{1-x-y} dz$. Moreover, by means of simplicity of notation, we have introduced
\begin{align}
{\pazocal{N}}^{\mu\nu\rho\sigma}=\textrm{tr}\Big[(\slashed{\ell}+\slashed{u}_{1}+m)\gamma^{\mu}
e^{\alpha\gamma^{5}}(\slashed{\ell}+\slashed{u}+m)\gamma^{\nu}
e^{\alpha\gamma^{5}}(\slashed{\ell}+\slashed{u}_{3}+m)\gamma^{\rho}
e^{\alpha\gamma^{5}}(\slashed{\ell}+\slashed{u}_{34}+m) \gamma^{\sigma}
e^{\alpha\gamma^{5}}\Big],
\label{eq:a14}
\end{align}
with the definitions $u_{i}=u-p_{i}$ and $u_{ij}=u-p_{i}-p_{j}$, and also
\begin{align}
\Delta&= x(x-1)p_{1}^2+(y+z)[(y+z)-1]p_{3}^2+y(y-1)p_{4}^2
+2[y(y-1)+yz] (p_{3}.p_{4})\nonumber\\
&+2x(z+y)(p_{3}.p_{1})+2 x y (p_{4}.p_{1})+m^2.
\label{eq:a13}
\end{align}

By means of definiteness, an important remark about the evaluation of \eqref{eq:a12} is in order because the integral should be regularized (it is logarithmically divergent).
Since we have chosen to use the dimensional regularization and the amplitude involves a $\gamma_5$ matrix, it is necessary to use the 't Hooft-Veltman rule in order to correctly define the $\gamma_5$ within the dimensional regularization  \cite{tHooft:1972tcz,Novotny:1994yx}.

The 't Hooft-Veltman method consists in the splitting of the $\omega$ dimensional spacetime into two parts: a 4-dimensional (physical) and a $(\omega-4)$-dimensional subspace
\begin{equation} \label{eq:31a}
\int d^{\omega}\ell\to\int d^{\omega}Q=\int d^{4}\ell\int d^{\omega-4}L.
\end{equation}
In this case, the internal momentum is expressed as below
\begin{equation} \label{eq:31}
\displaystyle{\not}Q=\displaystyle{\not}\ell+\displaystyle{\not}L=\left(\gamma^{0}\ell_{0}+...+\gamma^{3}\ell_{3}\right)
+\left(\gamma^{4}L_{4}+...+\gamma^{\omega-1}L_{\omega-1}\right),
\end{equation}
where we have denoted the internal momentum as $L$ for the remaining  $\omega-4$ components.

Within the 't Hooft-Veltman rule, the $\gamma_{5}$ algebra is written as
\begin{align}
\left\{ \gamma_{5},\gamma^{\mu}\right\}  & =0,\quad\mu=0,1,2,3\\
\left[\gamma_{5},\gamma^{\mu}\right] & =0,\quad\mu=4,...,\omega-1,
\end{align}
and all other familiar rules are still valid, including the algebra
\begin{equation}
\left\{ \gamma^{\mu},\gamma^{\nu}\right\} =2g^{\mu\nu},\quad\mu,\nu=0,1,...,\omega-1,
\end{equation}
with the metric tensor components $g_{\mu\nu}=\textrm{diag}\left(+1,-1,...,-1\right)$.
We observe that all the external momenta $p_{i}$ remain 4-dimensional,
this implies that $\displaystyle{\not}L^2 =-L^2$ and $\displaystyle{\not}L \displaystyle{\not}\ell+ \displaystyle{\not}\ell\displaystyle{\not}L=0$.

Consequently, applying the 't Hooft-Veltman rule in \eqref{eq:a14} i.e. $\ell\rightarrow \ell+L$, we find
\begin{align}
{\pazocal{N}}^{\mu\nu\rho\sigma}=\textrm{tr}\Big[(\slashed{Q}+\slashed{u}_{1}+m)\gamma^{\mu}
e^{\alpha\gamma^{5}}(\slashed{Q}+\slashed{u}+m)\gamma^{\nu}
e^{\alpha\gamma^{5}}(\slashed{Q}+\slashed{u}_{3}+m)\gamma^{\rho}
e^{\alpha\gamma^{5}}(\slashed{Q}+\slashed{u}_{34}+m) \gamma^{\sigma}
e^{\alpha\gamma^{5}}\Big]. \label{eq:a14a}
\end{align}
We then observe the presence of two possibly divergent terms, those proportional to $\ell^4$ and $L^4$. We shall focus our discussion on these terms, showing how they cancel and thus imply in the finiteness of the quantum effective action.

The numerator \eqref{eq:a14a} can be split into two parts according to the power of $m$ as below
\begin{align}
{\pazocal{N}}^{\mu\nu\rho\sigma}={\pazocal{N}}_{1}^{\mu\nu\rho\sigma}+{\pazocal{N}}_{2}^{\mu\nu\rho\sigma},
\label{eq:a15}
\end{align}
where, ${\pazocal{N}}_{1}^{\mu\nu\rho\sigma}$ and ${\pazocal{N}}_{2}^{\mu\nu\rho\sigma}$ include the odd and even powers of $m$, respectively.
Furthermore, we can write these terms in regard to the power of $m$ as the following
\begin{align}
\pazocal{N}_{1}^{\mu\nu\rho\sigma}&= \pazocal{N}^{\mu\nu\rho\sigma}_{(mu^3)}+\pazocal{N}^{\mu\nu\rho\sigma}_{(mu\ell^2)}+{\pazocal{N}}^{\mu\nu\rho\sigma}_{(m^3u)}
+ \pazocal{N}^{\mu\nu\rho\sigma}_{(muL^2)},\\
\pazocal{N}_{2}^{\mu\nu\rho\sigma}&= \pazocal{N}^{\mu\nu\rho\sigma}_{(\ell^4)}
+{\pazocal{N}}^{\mu\nu\rho\sigma}_{(\ell^2u^2)}
+\pazocal{N}^{\mu\nu\rho\sigma}_{(u^4)}+\pazocal{N}^{\mu\nu\rho\sigma}_{(\ell^2m^2)}
+\pazocal{N}^{\mu\nu\rho\sigma}_{(m^2u^2)}+\pazocal{N}^{\mu\nu\rho\sigma}_{(m^4)}
\nonumber\\
&+ \pazocal{N}^{\mu\nu\rho\sigma}_{(L^4)}+{\pazocal{N}}^{\mu\nu\rho\sigma}_{(L^2u^2)}
+{\pazocal{N}}^{\mu\nu\rho\sigma}_{(L^2\ell^2)}+\pazocal{N}^{\mu\nu\rho\sigma}_{(L^2m^2)}.
\label{eq:a16}
\end{align}
A first observation is that since every term in ${\pazocal{N}}_{1}^{\mu\nu\rho\sigma}$ includes a trace of an odd number of gamma matrices, we conclude that ${\pazocal{N}}_{1}^{\mu\nu\rho\sigma}=0$.
Thus, we now insert ${\pazocal{N}}_{2}^{\mu\nu\rho\sigma}$ into \eqref{eq:a12} and arrange it according to the power of the internal momentum $\ell$ as the following
\begin{align}
i\Lambda^{\mu\nu\rho\sigma}_{(1,a)}&=-\beta^{4}\Gamma(4) \int dX  \int \frac{d^{\omega}Q}{(2\pi)^{\omega}}
~\frac{
\pazocal{N}^{\mu\nu\rho\sigma}_{(m^4)}+\pazocal{N}^{\mu\nu\rho\sigma}_{(m^2u^2)}+{\pazocal{N}}^{\mu\nu\rho\sigma}_{(u^4)}
}{(\ell^2 -L^2-\Delta)^4},\\
i\Lambda^{\mu\nu\rho\sigma}_{(1,b)}&=-\beta^{4}\Gamma(4) \int dX  \int \frac{d^{\omega}Q}{(2\pi)^{\omega}}~\frac{
\pazocal{N}^{\mu\nu\rho\sigma}_{(\ell^2u^2)}+\pazocal{N}^{\mu\nu\rho\sigma}_{(\ell^2m^2)}
+ {\pazocal{N}}^{\mu\nu\rho\sigma}_{(L^2u^2)}
+\pazocal{N}^{\mu\nu\rho\sigma}_{(L^2m^2)}+{\pazocal{N}}^{\mu\nu\rho\sigma}_{(L^2\ell^2)}}{(\ell^2 -L^2-\Delta)^4},
\label{eq:a17-1}\\
i\Lambda^{\mu\nu\rho\sigma}_{(1,c)}&=-\beta^{4}\Gamma(4) \int dX  \int \frac{d^{\omega}Q}{(2\pi)^{\omega}} ~\frac{
\pazocal{N}^{\mu\nu\rho\sigma}_{(\ell^4)}+ \pazocal{N}^{\mu\nu\rho\sigma}_{(L^4)}
}{(\ell^2 -L^2-\Delta)^4}.
\label{eq:a17}
\end{align}
The explicit form of these terms in ${\pazocal{N}}_{2}^{\mu\nu\rho\sigma}$ can be found in the appendix A, eqs.~\eqref{eq:app1}-\eqref{eq:ap6}, and also the $L$ dependent terms in $\pazocal{N}$ are similar to those of $\ell$.

Now, we can show how the divergent contributions in \eqref{eq:a17} are cancelled independently.
As mentioned above, we have the following logarithmically divergent terms
\begin{align}
& \int \frac{d^{4}\ell }{(2\pi)^{4}} \frac{d^{\omega-4}L}{(2\pi)^{\omega-4}} \frac{1}{(\ell^2-L^2-\Delta)^4}  \textrm{tr}\Big[e^{4\alpha\gamma^{5}}
\slashed{\ell}\gamma^{\mu}\slashed{\ell} \gamma^{\nu}\slashed{\ell} \gamma^{\rho}\slashed{\ell} \gamma^{\sigma}\Big]  \label{eq:a23l}\\
&\int \frac{d^{4}\ell }{(2\pi)^{4}} \frac{d^{\omega-4}L}{(2\pi)^{\omega-4}} \frac{1}{(\ell^2-L^2-\Delta)^4}   \textrm{tr} \Big[\slashed{L}\gamma^{\mu}\slashed{L}\gamma^{\nu}\slashed{L}\gamma^{\rho}
\slashed{L}\gamma^{\sigma}\Big], \label{eq:a23L}
\end{align}
we notice that both terms have similar tensor structure (due to the trace operation).
Thus, the tensor and regularized structure is nearly common for the above $\ell^4$ and $L^4$ terms and we can use it to apply our considerations.
Hence, we can evaluate straightforwardly the momentum integration with help of \eqref{eq:app1} and use of the identity $Q_\delta Q_\tau Q_\xi Q_\eta \to \frac{Q^4}{\omega(\omega+2)} \left( g_{\delta\tau}g_{\xi\eta}+g_{\delta\xi}g_{\tau\eta}
+g_{\delta\eta}g_{\tau\xi}\right)  $, and show that both contributions are proportional to $\Gamma\left(\frac{\epsilon}{2}\right)$, where $\epsilon=4-\omega \to 0^+$, and also to the tensor structure
\begin{align}
\widetilde{{\pazocal{N}}}^{\mu\nu\rho\sigma}&= \frac{1}{4}\Big(g_{\delta\tau}g_{\xi\eta}+g_{\delta\xi}g_{\tau\eta}
+g_{\delta\eta}g_{\tau\xi}\Big) \textrm{Tr}\Big[e^{4\alpha\gamma^{5}}\gamma^{\delta}\gamma^{\mu}
\gamma^{\tau}\gamma^{\nu}
\gamma^{\xi}\gamma^{\rho}
\gamma^{\eta} \gamma^{\sigma}
\Big], \cr
&= \frac{1}{4}\Big(g_{\delta\tau}g_{\xi\eta}+g_{\delta\xi}g_{\tau\eta}
+g_{\delta\eta}g_{\tau\xi}\Big) \textrm{Tr}\Big[\Big(\cosh(4\alpha)+\sinh(4\alpha)\gamma^{5}\Big)\gamma^{\delta}\gamma^{\mu}
\gamma^{\tau}\gamma^{\nu}\gamma^{\xi}\gamma^{\rho}\gamma^{\eta}\gamma^{\sigma}\Big],
\label{eq:a24}
\end{align}
while the $L$ contribution in eq.~\eqref{eq:a23L} is obtained from \eqref{eq:a24} as $\alpha=0$.
Moreover, we notice that the terms including $\cosh(4\alpha)$ and $\sinh(4\alpha)$ refer to the parity-preserving and parity-violating contributions, respectively. However, in the usual QED, there is only terms with $\alpha\to 0$, so that $\beta \to g_v$.

Finally, making use of this result and by taking into account all the 24 (tensor and momentum) permutations of \eqref{eq:a17}, and also performing the relevant traces through FeynCalc program, in both parity-preserving and parity-violating pieces, we find that
\begin{equation}
\Lambda^{\mu\nu\rho\sigma}_{\tiny\mbox{total}}\Big|_{div.}=\sum\limits_{i=1}^{24}\Lambda^{\mu\nu\rho\sigma}_{(i,c)}\Big|_{div.}=0,
\label{eq:a26}
\end{equation}
which is in accordance to the result of the usual QED.
Thus, we arrive at the same finite result for the total amplitude as in equation \eqref{eq:a11}.

After showing the UV-finiteness of the total amplitude, we are now ready to obtain the generalized expression for the Euler-Heisenberg effective Lagrangian.
Thus, by making use of the standard Feynman integrals, we arrive at the following expressions for the finite part
\begin{align}
\Lambda^{\mu\nu\rho\sigma}_{(1,a)}&=-\frac{\beta^{4}}{16\pi^{2}}\int dX ~\frac{1}{\Delta^{2}}~\Big[
\pazocal{N}^{\mu\nu\rho\sigma}_{(m^4)}+{\pazocal{N}}^{\mu\nu\rho\sigma}_{(m^2u^2)}+
{\pazocal{N}}^{\mu\nu\rho\sigma}_{(u^4)}
\Big],\\
\Lambda^{\mu\nu\rho\sigma}_{(1,b)}&=\frac{\beta^{4}}{16\pi^{2}}\int dX ~\frac{1}{\Delta}~\Big[
\widetilde{{\pazocal{N}}}^{\mu\nu\rho\sigma}_{(u^2)}+\widetilde{{\pazocal{N}}}^{\mu\nu\rho\sigma}_{(m^2 )}
+
\widehat{{\pazocal{N}}}^{\mu\nu\rho\sigma}_{(u^2)}+\widehat{{\pazocal{N}}}^{\mu\nu\rho\sigma}_{(m^2 )}
+\widehat{{\pazocal{N}}}^{\mu\nu\rho\sigma}_{(0)}\Big],
\label{eq:a18}
\end{align}
the quantities with tilde and hat in \eqref{eq:a18} indicate the integrated expressions of \eqref{eq:a17-1} over the internal momenta $\ell$ and $L$.

Assuming on-shell photons, $p_{i}^2=0$, the expression $\Delta$ in
\eqref{eq:a13} changes to
\begin{align}
\Delta=m^2(1+\xi),
\label{eq:a19}
\end{align}
where
\begin{equation}
\xi=2y(y-1+z) (\frac{p_{3}.p_{4}}{m^2})+2x(z+y)(\frac{p_{3}.p_{1}}{m^2})+2 x y (\frac{p_{4}.p_{1}}{m^2}).
\label{eq:a20}
\end{equation}
Now, in the low-energy limit the photon energies are small compared  to the fermionic mass $m$, i.e. $p_i.p_j \ll m^2$.
Under these considerations, we obtain
\begin{align}
\Lambda^{\mu\nu\rho\sigma}_{(1,a)}&=-\frac{\beta^{4}}{16\pi^{2}}\frac{1}{m^{4}}\int dX ~\Big[
{\pazocal{N}}^{\mu\nu\rho\sigma}_{(m^4)}+{\pazocal{N}}^{\mu\nu\rho\sigma}_{(m^2u^2)}+{\pazocal{N}}^{\mu\nu\rho\sigma}_{(u^4)}
\Big]\Big[1-2\xi + 3\xi^2\Big] \label{eq:a21},\\
\Lambda^{\mu\nu\rho\sigma}_{(1,b)}&= \frac{\beta^{4}}{16\pi^{2}}\frac{1}{m^{2}}\int dX ~\Big[
\widetilde{{\pazocal{N}}}^{\mu\nu\rho\sigma}_{(u^2)}+\widetilde{{\pazocal{N}}}^{\mu\nu\rho\sigma}_{(m^2 )}
+\widehat{{\pazocal{N}}}^{\mu\nu\rho\sigma}_{(u^2)}+\widehat{{\pazocal{N}}}^{\mu\nu\rho\sigma}_{(m^2 )}
+\widehat{{\pazocal{N}}}^{\mu\nu\rho\sigma}_{(0)}
\Big]\Big[1-\xi + \xi^2 \Big].
 \label{eq:a22}
\end{align}
As we have previously mentioned, we observe that $\Lambda^{\mu\nu\rho\sigma}_{(1,a)}$ and $\Lambda^{\mu\nu\rho\sigma}_{(1,b)}$ are UV finite.

To determine the generalized Euler-Heisenberg effective action, we should concentrate in examining the term which includes four momenta of the external photons, i.e. ${\pazocal{N}}^{\mu\nu\rho\sigma}_{(u^4)}$ in \eqref{eq:a21}, and also discarding the $\xi \ll 1$ parts. Thus, using the explicit form of ${\pazocal{N}}^{\mu\nu\rho\sigma}_{(u^4)}$ in \eqref{eq:app-5}, we find
\begin{align}
\Lambda^{\mu\nu\rho\sigma}_{(1,a)}\Big|_{u^{4}}^{\xi^{0}}=-\frac{\beta^{4}}{16\pi^{2}m^{4}}
\textrm{Tr}\Big([\cosh(4\alpha)+\sinh(4\alpha)\gamma^{5}]\gamma^{\delta}\gamma^{\mu}
\gamma^{\tau}\gamma^{\nu}
\gamma^{\xi}\gamma^{\rho}
\gamma^{\eta} \gamma^{\sigma}
\Big)\int dX ~
u_{1\delta}u_{\tau}u_{3\xi}u_{34\eta}.
\label{eq:a27}
\end{align}
Now, by making use of eq.~\eqref{eq:a3} to return to the $(g_v,g_a)$ couplings, and separating eq.~\eqref{eq:a27} into the parity-conserving (P.C) and parity-violating (P.V) contributions, we have
\begin{align}
\Big(\Lambda^{\mu\nu\rho\sigma}_{(1,a)}\Big|_{u^{4}}^{\xi^{0}}\Big)_{P.C}&=
-\frac{(g_{v}^{4}+g_{a}^{4}+6g_{v}^{2}g_{a}^{2})}{16\pi^{2}m^{4}} ~\textrm{Tr}\Big(\gamma^{\delta}\gamma^{\mu}
\gamma^{\tau}\gamma^{\nu}
\gamma^{\xi}\gamma^{\rho}
\gamma^{\eta} \gamma^{\sigma}
\Big)\int dX ~
u_{1\delta}u_{\tau}u_{3\xi}u_{34\eta},
 \label{eq:a28}\\
\Big(\Lambda^{\mu\nu\rho\sigma}_{(1,a)}\Big|_{u^{4}}^{\xi^{0}}\Big)_{P.V}&=-\frac{4(g_{v}^{3}g_{a}+g_{v}g_{a}^{3})}
{16\pi^{2}m^{4}}~\textrm{Tr}\Big(\gamma^{5}\gamma^{\delta}\gamma^{\mu}
\gamma^{\tau}\gamma^{\nu}
\gamma^{\xi}\gamma^{\rho}
\gamma^{\eta} \gamma^{\sigma}
\Big)\int dX ~
u_{1\delta}u_{\tau}u_{3\xi}u_{34\eta}.
\label{eq:a29}
\end{align}
A first comment about these expressions is that in the limit $g_{a}\rightarrow 0$, the P.V contribution vanishes and the value of P.C reduces to the ordinary result.
Furthermore, we observe that both contributions, P.C and P.V, are totally symmetric under the exchange of $g_{v}\leftrightarrow g_{a}$, corroborating our arguments of the presence of this symmetry for the  $n=4$ point function.
However, this symmetry property is in disagreement with the behaviour of the results found in ref. \cite{Yamashita:2017}.
But we should emphasize that our symmetry arguments are general, since the parametrization \eqref{eq:a3} is independent of the perturbative analysis (that reflects the generality of our arguments).

After performing the remaining traces and taking the integral over the Feynman parameters in \eqref{eq:a28} and \eqref{eq:a29} through FeynCalc program, we should apply all of the 24 permutations
\begin{align}
\Big(\Lambda^{\mu\nu\rho\sigma}\Big|_{u^{4}}^{\xi^{0}}\Big)_{P.C}
=\sum\limits_{i=1}^{24}\Big(\Lambda^{\mu\nu\rho\sigma}_{(i,a)}\Big|_{u^{4}}^{\xi^{0}}\Big)_{P.C},~~~~~~
\Big(\Lambda^{\mu\nu\rho\sigma}\Big|_{u^{4}}^{\xi^{0}}\Big)_{P.V}
=\sum\limits_{i=1}^{24}\Big(\Lambda^{\mu\nu\rho\sigma}_{(i,a)}\Big|_{u^{4}}^{\xi^{0}}\Big)_{P.V},
\label{eq:a30}
\end{align}
which finally lead to the following expression of the total parity-preserving contribution
\begin{align}
\Big(\Lambda^{\mu\nu\rho\sigma}\Big|_{u^{4}}^{\xi^{0}}\Big)_{P.C}&=
\frac{(g_{v}^{4}+g_{a}^{4}+6g_{v}^{2}g_{a}^{2})}{60\pi^{2}m^{4}}
\cr
&\times \Bigg[\frac{7}{3}\Big(p_1^{\rho } p_3^{\sigma } p_4^{\nu } p_3^{\mu }+p_1^{\rho } p_{4}^{\nu } p_{4}^{\sigma } p_{3}^{\mu }-p_{4}^{\nu } p_{1}^{\rho }p_{1}^{\sigma }p_{3}^{\mu } +p_{1}^{\nu } p_{3}^{\sigma }p_{4}^{\mu } p_{3}^{\rho }
+ p_{1}^{\nu } p_{4}^{\mu } p_{4}^{\sigma } p_{3}^{\rho }-p_{4}^{\mu } p_{1}^{\nu } p_{1}^{\sigma }  p_{3}^{\rho }\Big) \cr
& +\Big(p_{1}^{\nu} p_{1}^{\rho} p_{4}^{\sigma} p_{3}^{\mu}-p_{4}^{\nu} p_{3}^{\rho}p_{1}^{\sigma }p_{3}^{\mu} -p_{1}^{\nu } p_{3}^{\rho }p_{4}^{\sigma }p_{3}^{\mu }-p_{4}^{\mu } p_{4}^{\nu } p_{1}^{\sigma }  p_{3}^{\rho}
+p_{1}^{\nu } p_{1}^{\rho } p_{4}^{\mu } p_{3}^{\sigma }- p_{4}^{\mu }p_{4}^{\nu } p_{1}^{\rho }  p_{3}^{\sigma}\Big)\Bigg],
\label{eq:a31}
\end{align}
and also to the total parity-violating contribution
\begin{align}
\Big(\Lambda^{\mu\nu\rho\sigma}\Big|_{u^{4}}^{\xi^{0}}\Big)_{P.V}=&-\frac{i(g_{v}^{3}g_{a}+g_{v}g_{a}^{3})}
{24\pi^{2}m^{4}}
\Big[ p_{1}^{\sigma }  p_{3}^{\mu } \epsilon ^{\nu   p_{3}  p_{4} \rho }+  p_{1}^{\sigma }  p_{4}^{\mu } \epsilon ^{\nu   p_{3}  p_{4} \rho }-  p_{1}^{\rho }  p_{4}^{\sigma } \epsilon ^{\mu  \nu   p_{1}  p_{3}}
+  p_{1}^{\rho }  p_{4}^{\mu } \epsilon ^{\nu   p_{1}  p_{3} \sigma }
\cr
&+  p_{1}^{\rho }  p_{4}^{\mu } \epsilon ^{\nu   p_{3}  p_{4} \sigma }-  p_{1}^{\nu }  p_{3}^{\sigma } \epsilon ^{\mu   p_{1}  p_{4} \rho }
 +p_{1}^{\nu }  p_{3}^{\mu } \epsilon ^{ p_{1}  p_{4} \rho  \sigma }-  p_{1}^{\nu }  p_{3}^{\mu } \epsilon ^{ p_{3}  p_{4} \rho  \sigma }
\cr
&+  p_{3}^{\sigma }  p_{4}^{\nu } \epsilon ^{\mu   p_{1}  p_{4} \rho }
+  p_{3}^{\rho }  p_{4}^{\sigma } \epsilon ^{\mu  \nu   p_{1}  p_{3}}+  p_{3}^{\rho }  p_{4}^{\nu } \epsilon ^{\mu   p_{1}  p_{3} \sigma }+  p_{3}^{\rho }  p_{4}^{\nu } \epsilon ^{\mu   p_{1}  p_{4} \sigma }\Big].
\label{eq:a32}
\end{align}
One should note that in eq.~\eqref{eq:a32} the coefficient $i$ arises from the trace of $\gamma^{5}$ with the gamma matrices.
Now that we have evaluated the one-loop effective action for the generalized parity-violating coupling, we are in a position to determine the effective couplings $\lambda_i$'s by comparing these results to those from Sec. 3.


\section{Determination of the effective couplings $(\lambda_1,\lambda_2,\lambda_3)$}
\label{sec6}


At this stage, we can determine the value of three coupling constants appearing in the effective Lagrangian \eqref{eq:b12}
\begin{align}
\pazocal{L}_{ext.}^{^{E.H}}=\lambda_{1}\pazocal{F}^{2}+\lambda_{2}\pazocal{G}^{2}+\lambda_{3}\pazocal{F}\pazocal{G}.
\label{eq:effective-L}
\end{align}
This can be achieved by matching the results found in eqs.\eqref{eq:b17} and \eqref{eq:a31} for the parity-preserving piece
\begin{align}
\Big(\Xi^{\mu\nu\rho\sigma}\Big)_{P.C}=\Big(\Lambda^{\mu\nu\rho\sigma}\Big|_{u^{4}}^{\xi^{0}}\Big)_{P.C},
\label{eq:a33a}
\end{align}
as well as eqs.\eqref{eq:b21} and \eqref{eq:a32} for the parity-violating part
\begin{align}
\Big(\Xi^{\mu\nu\rho\sigma}\Big)_{P.V}=\Big(\Lambda^{\mu\nu\rho\sigma}\Big|_{u^{4}}^{\xi^{0}}\Big)_{P.V}.
\label{eq:a33b}
\end{align}
After some algebraic manipulations, we obtain the values of $(\lambda_1,\lambda_2)$ from eq.\eqref{eq:a33a}, while \eqref{eq:a33b} gives us the value of $\lambda_3$. These expressions are summarized as
\begin{align}
\lambda_1&=\frac{1}{512\pi^{2}m^4}\left[\frac{16}{45}g_{v}^4+\frac{32}{15}g_{v}^2 g_{a}^2+\frac{16}{45}g_{a}^4\right], \cr
\lambda_2&=\frac{1}{512\pi^{2}m^4}\left[\frac{28}{45}g_{v}^4+\frac{56}{15}g_{v}^2 g_{a}^2+\frac{28}{45}g_{a}^4\right],\cr
\lambda_3&=-\frac{i}{768\pi^2 m^4}\left[g_{v}^{3}g_{a}+g_{v}g_{a}^{3}\right].
\label{eq:a34}
\end{align}
Since the one-loop $4$-point function is symmetric under the change of $g_{v}\leftrightarrow g_{a}$, it was expected that the effective couplings $\lambda_i$ present a similar behaviour, as verified in \eqref{eq:a34}.
Here, as we have mentioned above, we observe again that the values of the coupling constants are different from those obtained in the ref.\cite{Yamashita:2017}, that do not display such a symmetric behaviour.

To clarify this issue, let us consider two cases in the interacting Lagrangian \eqref{eq:a2}, or equivalently in the effective couplings \eqref{eq:a34}:
\begin{align}
\lambda_{1}^{\rm EH}=\frac{g_{v}^4}{1440\pi^{2}m^4},~~~~\lambda_{2}^{\rm EH}=\frac{7g_{v}^4}{5760\pi^{2}m^4},~~~~\lambda_{3}=0,
\label{eq:a35}
\end{align}
that correspond to the two limiting cases:  $g_{a}=0$, $g_{v}\neq 0$ and  $g_{v}=0$, $g_{a}\neq 0$.

 Both cases, in contrast to \cite{Yamashita:2017}, lead to the same result for the three effective coupling constants as the following which coincides exactly with effective couplings of an ordinary parity-conserving Euler-Heisenberg Lagrangian \cite{Dittrich:2000zu,Preucil:2017}.

Therefore, we realize that considering the pure-vector interaction (even-parity) or the pure-axial interaction (odd-parity) of the gauge field with fermionic matter leads to an identical effective action, which is parity-preserving.
In the other words, the interaction term with a distinct behaviour under parity produces a parity-conserving effective action.
While the case  $g_{v}\neq 0$, $g_{a}\neq 0$ yields us a generalized effective Euler-Heisenberg action with parity-conserving and parity-violating terms \eqref{eq:a34}.

As the last case, we consider the V+A and V-A interactions, which corresponds to $g_{a}=\pm g_{v}$ , respectively,
\begin{equation}
{\pazocal{L}}_{\tiny\mbox{V+A}}=-g_{v}\bar\psi\gamma^{\mu}(1+\gamma^{5})A_{\mu}\psi,\,\,\
{\pazocal{L}}_{\tiny\mbox{V-A}}=-g_{v}\bar\psi\gamma^{\mu}(1-\gamma^{5})A_{\mu}\psi,
\label{eq:a36}
\end{equation}
and find the effective coupling constants as follows
\begin{align}
\lambda_{1}\Big|_{g_{a}=\pm g_{v}}=\frac{g_{v}^4}{180\pi^{2}m^4},~~~~\lambda_{2}\Big|_{g_{a}=\pm g_{v}}=\frac{7g_{v}^4}{720\pi^{2}m^4},~~~~\lambda_{3}\Big|_{g_{a}=\pm g_{v}}=\mp\frac{ig_{v}^4}{384\pi^2 m^4}. \label{VA}
\end{align}
Naturally, as we expected, the values of the constants $\lambda_{i}$'s \eqref{VA} are also in disagreement with the results obtained in \cite{Yamashita:2017}.

\section{$\gamma \gamma \to \gamma \gamma$ differential cross section}
\label{sec7}

In order to examine the phenomenology of the effective action \eqref{eq:effective-L}, with the couplings \eqref{eq:a34}, it is interesting to calculate the differential cross section for the light by light scattering.
We would like to recall that, although the following analysis should contain terms from the interference of QED with the standard model's lowest order parity-violating terms, we have already discussed that these standard model's contribution are negligible, and shall not be considered here.

Therefore, to start the analysis of the differential cross section for the light by light scattering, we first consider the tensorial structure of the total amplitude in \eqref{eq:b22} as the following
\begin{align}
\Xi^{\mu\nu\rho\sigma}=\Big(\Xi^{\mu\nu\rho\sigma}\Big)_{P.C}+\Big(\Xi^{\mu\nu\rho\sigma}\Big)_{P.V}.
\label{eq:a37}
\end{align}
As a first check, we verify the Ward identity by inserting eqs.~\eqref{eq:b17} and \eqref{eq:b21} back into \eqref{eq:a37}, and then contracting the resulting expression with the external momenta, which give the following
\begin{equation}
  p_{1\mu}\Xi^{\mu\nu\rho\sigma}=p_{2\sigma}\Xi^{\mu\nu\rho\sigma}= p_{3\nu}\Xi^{\mu\nu\rho\sigma}=p_{4\rho}\Xi^{\mu \nu\rho\sigma}=0.
  \label{eq:a38}
\end{equation}
This result is expected since the effective Lagrangian \eqref{eq:effective-L} is constructed from the gauge invariant quantities $\pazocal{F}$ and $\pazocal{G}$.
 The unpolarized differential cross section is proportional to the average of the absolute square of the
total amplitude as
$\frac{1}{4}\sum |\pazocal{M} |^2$ that includes the following sum over the photon polarization states
\begin{equation}
  \sum\limits_{\lambda=1}^{2} \varepsilon^{\ast}_{\mu}(p,\lambda) \varepsilon_{\nu}(p,\lambda) = - g_{\mu \nu} + \mathbb{L}_{\mu
  \nu},
\end{equation}
where $\mathbb{L}_{\mu \nu}$ is a longitudinal part which contraction with $\Xi^{\mu\nu\rho\sigma}$ vanishes, due to the Ward identity \eqref{eq:a38}.
In the center of mass frame (CM), we define $\omega$ as the photon energy and $\theta$ as the departure
angle of the final state photons.
Hence, the unpolarized differential cross section is given by
\begin{equation}
  \frac{d \sigma}{d \Omega} \Big|_{CM}= \Big(278 g^8_a + 3561 g^6_a g^2_v + 11014  g^4_a
  g^4_v + 3561 g^2_a g^6_v + 278 g^8_v\Big) \left(\frac{\omega^6}{m^8}\right)
  \frac{ \left[7 + \cos(2\theta)\right]^2}{66355200 \pi^6 }. \label{diff_CS}
\end{equation}
It is worth noticing that the presence of the CP-violating term $\lambda_{3}\pazocal{F}\pazocal{G}$ does not change the angular distribution of the differential cross section of the QED description.
Rather, its contribution is only as a numerical factor; therefore, an enhancement in regard of the usual value measured in the related cross section could be a signal of the presence of parity violation in QED.

Here, we mention that the obtained result \eqref{diff_CS} is also symmetric under the exchange of $g_a \leftrightarrow g_v$. Moreover,
in the pure-vector and pure-axial coupling limits, we arrive at
\begin{equation}
  \lim\limits_{g_{a}\rightarrow 0}\frac{d \sigma}{d \Omega}\Big|_{CM} =
   \lim\limits_{g_{v}\rightarrow 0}\frac{d \sigma}{d \Omega}\Big|_{CM} = \frac{139 g^8_v }{33177600 \pi^6} \left(\frac{\omega^6}{m^8}\right)
  \left[7+\cos(2\theta)\right]^2.
\end{equation}
We observe that these limiting cases yield us the same result, which coincides exactly with the standard differential cross section for the photon-photon scattering in the usual QED \cite{heisenberg-euler,Berestetsky:1982}, as expected.

\section{Conclusion}
\label{conc}

In this paper, we have perturbatively examined the Euler-Heisenberg effective action in the presence of an axial coupling of the gauge field with the fermionic matter (that might be ascribed to new physics, for instance, dark matter particles \cite{Fan:2017sxk}).
These novel coupling effects are responsible for generating a CP-violating term $\pazocal{F}\pazocal{G}$ in the effective action.
We emphasize that the axial anomalous term added is not an acceptable extension of QED and thus the induced CP-violating
term in the usual Euler-Heisenberg model as an effective Lagrangian can be obtained only from this anomalous theory, and not from any fundamental field theory.
One important aspect of our analysis is in regard of the regularization to the Feynman amplitude. Since the amplitude of the process $\gamma\gamma\to\gamma\gamma$ is divergent, and we have the presence of an axial coupling in terms of the $\gamma_5$ matrix, we presented a detailed analysis of the algebraic manipulations using the 't Hooft-Veltman rule.
We have shown the presence of two possibly divergent terms, and discussed how their contributions are cancelled when 24 permutations in the box diagram are considered.

Another point is related to a hidden symmetry of the vector and axial couplings, which was observed in terms of a proper parametrization.
The piece of the 4-point function contributing to the extended Euler-Heisenberg effective Lagrangian is written in terms of the parametrized couplings $\beta^{4}e^{4\alpha\gamma^{5}}=(g_{v}^{4}+g_{a}^{4}+6g_{v}^{2}g_{a}^{2})+4(g_{v}^{3}g_{a}+g_{v}g_{a}^{3})\gamma^{5} $, which allows us to verify that it is symmetric under the change of $g_{v}\leftrightarrow g_{a}$.
Furthermore, this observation also allowed us to conclude that the usual Euler-Heisenberg effective action, which is parity-preserving, can be generated whether by the pure-vector interaction (even-parity) or the pure-axial interaction (odd-parity), see eq.\eqref{eq:a35}.
It is important to remark that our results are in contrast to those of ref.~\cite{Yamashita:2017}.
But we believe that, although following different approaches, a crucial point showing the correctness of our analysis is the presence of the aforementioned symmetry in the $\lambda_i$ coefficients of the extended Euler-Heisenberg effective Lagrangian, see eq.\eqref{eq:a34}.
Actually, this hidden symmetry of the box graph under $g_{v}\leftrightarrow g_{a}$ is absent in the results  of ref.~\cite{Yamashita:2017}.

We proceeded with a phenomenological application of the generalized effective action, considering the effects of the CP-violating term (as a possible new physics phenomenon), to compute the differential cross section related to the photon-photon scattering.
As a result, we found that the angular distribution of the differential cross section in the presence of CP-violating term does not modify, except some changes in the numerical coefficients. Hence, the effects of the parity-violating term could be measured in terms of the enhanced value of the cross section.
It is worth to mention that the interference of the standard model
CP-violating contribution (from the  phase in  the CKM matrix  or from the $\theta$-term in QCD),
with the one in the present analysis, are by far smaller \cite{Millo:2008ug,Gorghetto:2021luj} and thus not considered.

\subsection*{Acknowledgements}
We would like to express our especial thanks to M. Chaichian for his valuable comments and many illuminating discussions. Also, we thank M.M. Sheikh-Jabbari and M. Mohammadi for their insightful comments and suggestions.
R.B. acknowledges partial support from Conselho
Nacional de Desenvolvimento Cient\'ifico e Tecnol\'ogico (CNPq Projects No. 305427/2019-9 and No. 421886/2018-8) and Funda\c{c}\~ao de
Amparo \`a Pesquisa do Estado de Minas Gerais (FAPEMIG Project No. APQ-01142-17).\\\\
\appendix
\section{Appendix:~The explicit form of $ {\pazocal{N}}_{2}^{\mu\nu\rho\sigma}$}\setcounter{equation}{0}\noindent
\label{apendixA}

 Here, we present the explicit form of the six terms appeared in the relation \eqref{eq:a16}
  \begin{align*}
 \pazocal{N}_{2}^{\mu\nu\rho\sigma}&= \pazocal{N}^{\mu\nu\rho\sigma}_{(\ell^4)}
+{\pazocal{N}}^{\mu\nu\rho\sigma}_{(\ell^2u^2)}
+\pazocal{N}^{\mu\nu\rho\sigma}_{(u^4)}+\pazocal{N}^{\mu\nu\rho\sigma}_{(m^2\ell^2)}
+\pazocal{N}^{\mu\nu\rho\sigma}_{(m^2u^2)}+\pazocal{N}^{\mu\nu\rho\sigma}_{(m^4)}
\nonumber\\
&+ \pazocal{N}^{\mu\nu\rho\sigma}_{(L^4)}+{\pazocal{N}}^{\mu\nu\rho\sigma}_{(L^2u^2)}
+{\pazocal{N}}^{\mu\nu\rho\sigma}_{(L^2\ell^2)}+\pazocal{N}^{\mu\nu\rho\sigma}_{(m^2L^2)},
  \end{align*}
 as the following
 \begin{align}
{\pazocal{N}}^{\mu\nu\rho\sigma}_{(\ell^4)}&= \ell_{\delta}\ell_{\tau}\ell_{\xi}\ell_{\eta} \textrm{Tr}\Big[e^{4\alpha\gamma^{5}}\gamma^{\delta}\gamma^{\mu}
\gamma^{\tau}\gamma^{\nu}
\gamma^{\xi}\gamma^{\rho}
\gamma^{\eta} \gamma^{\sigma}
\Big]
\label{eq:app1},
\\
{\pazocal{N}}^{\mu\nu\rho\sigma}_{(\ell^2u^2)}&=\Big[
{u}_{3\xi}{u}_{34\eta} \ell_{\delta}\ell_{\tau}+
u_{\tau} {u}_{34\eta}\ell_{\delta} \ell_{\xi}+
  \ell_{\delta}\ell_{\eta}u_{\tau} u_{3\xi}+
\ell_{\tau}\ell_{\xi}u_{1\delta}u_{34\eta}+
\ell_{\tau}\ell_{\eta}u_{1\delta}u_{3\xi}+
\ell_{\xi}\ell_{\eta}u_{1\delta}u_{\tau}
 \Big]\cr
  &\times \textrm{Tr}\Big[e^{4\alpha\gamma^{5}}\gamma^{\delta}\gamma^{\mu}
\gamma^{\tau}\gamma^{\nu}
\gamma^{\xi}\gamma^{\rho}
\gamma^{\eta} \gamma^{\sigma}
\Big],
\\
{\pazocal{N}}^{\mu\nu\rho\sigma}_{(m^2\ell^2)}&=m^2 \ell_{\delta}\ell_{\tau}\Big[ \textrm{Tr}\big(e^{2\alpha\gamma^{5}}\gamma^{\delta}\gamma^{\mu}
\gamma^{\tau}\gamma^{\nu}
\gamma^{\rho}
 \gamma^{\sigma}
\big)+\textrm{Tr}\big(\gamma^{\delta}\gamma^{\mu}
\gamma^{\nu}\gamma^{\tau}
\gamma^{\rho}
 \gamma^{\sigma}
\big)+\textrm{Tr}\big(e^{2\alpha\gamma^{5}}\gamma^{\delta}\gamma^{\mu}
\gamma^{\nu}
\gamma^{\rho}\gamma^{\tau}
 \gamma^{\sigma}
\big)\cr
&+ \textrm{Tr}\big(e^{-2\alpha\gamma^{5}}\gamma^{\mu}\gamma^{\delta}
\gamma^{\nu}\gamma^{\tau}
\gamma^{\rho}
 \gamma^{\sigma}\big)
 +\textrm{Tr}\big(\gamma^{\mu}\gamma^{\delta}
\gamma^{\nu}
\gamma^{\rho}\gamma^{\tau}
 \gamma^{\sigma}
\big)
+\textrm{Tr}\big(e^{2\alpha\gamma^{5}}\gamma^{\mu}
\gamma^{\nu}\gamma^{\delta}
\gamma^{\rho}
\gamma^{\tau} \gamma^{\sigma}
\big)\Big],
\\
{\pazocal{N}}^{\mu\nu\rho\sigma}_{(L^4)}&= L_{\delta}L_{\tau}L_{\xi}L_{\eta} \textrm{Tr}\Big[\gamma^{\delta}\gamma^{\mu}
\gamma^{\tau}\gamma^{\nu}
\gamma^{\xi}\gamma^{\rho}
\gamma^{\eta} \gamma^{\sigma}
\Big]
,
\\
 {\pazocal{N}}^{\mu\nu\rho\sigma}_{(m^2L^2)}&=m^2 L_{\delta}L_{\tau}\Big[ \textrm{Tr}\big(\gamma^{\delta}\gamma^{\mu}
\gamma^{\tau}\gamma^{\nu}
\gamma^{\rho}
 \gamma^{\sigma}
\big)+\textrm{Tr}\big(\gamma^{\delta}\gamma^{\mu}
\gamma^{\nu}\gamma^{\tau}
\gamma^{\rho}
 \gamma^{\sigma}
\big)+\textrm{Tr}\big(\gamma^{\delta}\gamma^{\mu}
\gamma^{\nu}
\gamma^{\rho}\gamma^{\tau}
 \gamma^{\sigma}
\big)\cr
&+ \textrm{Tr}\big(\gamma^{\mu}\gamma^{\delta}
\gamma^{\nu}\gamma^{\tau}
\gamma^{\rho}
 \gamma^{\sigma}\big)
 +\textrm{Tr}\big(\gamma^{\mu}\gamma^{\delta}
\gamma^{\nu}
\gamma^{\rho}\gamma^{\tau}
 \gamma^{\sigma}
\big)
+\textrm{Tr}\big(\gamma^{\mu}
\gamma^{\nu}\gamma^{\delta}
\gamma^{\rho}
\gamma^{\tau} \gamma^{\sigma}
\big)\Big], \\
{\pazocal{N}}^{\mu\nu\rho\sigma}_{(L^2\ell^2)}&=\ell_{\delta}\ell_{\tau}L_{\xi}L_{\eta}\Big[\textrm{Tr}\big(e^{2\alpha\gamma^5}
\gamma^{\delta}\gamma^{\mu}\gamma^{\xi}\gamma^{\nu}\gamma^{\eta}\gamma^{\rho}\gamma^{\tau}\gamma^{\sigma}\big)
+
\textrm{Tr}\big(e^{2\alpha\gamma^5}
\gamma^{\xi}\gamma^{\mu}\gamma^{\eta}\gamma^{\nu}\gamma^{\delta}\gamma^{\rho}\gamma^{\tau}\gamma^{\sigma}\big)\nonumber\\
&+
\textrm{Tr}\big(e^{2\alpha\gamma^5}
\gamma^{\delta}\gamma^{\mu}\gamma^{\tau}\gamma^{\nu}\gamma^{\xi}\gamma^{\rho}\gamma^{\eta}\gamma^{\sigma}\big)
+
\textrm{Tr}\big(e^{-2\alpha\gamma^5}
\gamma^{\xi}\gamma^{\mu}\gamma^{\delta}\gamma^{\nu}\gamma^{\tau}\gamma^{\rho}\gamma^{\eta}\gamma^{\sigma}\big)
\nonumber\\
&+
\textrm{Tr}\big(
\gamma^{\delta}\gamma^{\mu}\gamma^{\xi}\gamma^{\nu}\gamma^{\tau}\gamma^{\rho}\gamma^{\eta}\gamma^{\sigma}\big)
+
\textrm{Tr}\big(
\gamma^{\xi}\gamma^{\mu}\gamma^{\delta}\gamma^{\nu}\gamma^{\eta}\gamma^{\rho}\gamma^{\tau}\gamma^{\sigma}\big)
\Big],
\end{align}
\begin{align}
  {\pazocal{N}}^{\mu\nu\rho\sigma}_{(L^2u^2)}&=L_{\xi}L_{\eta}\Big[u_{1\delta}u_{34\tau}\textrm{Tr}\big(e^{2\alpha\gamma^5}
\gamma^{\delta}\gamma^{\mu}\gamma^{\xi}\gamma^{\nu}\gamma^{\eta}\gamma^{\rho}\gamma^{\tau}\gamma^{\sigma}\big)
+
u_{3\delta}u_{34\tau}\textrm{Tr}\big(e^{2\alpha\gamma^5}
\gamma^{\xi}\gamma^{\mu}\gamma^{\eta}\gamma^{\nu}\gamma^{\delta}\gamma^{\rho}\gamma^{\tau}\gamma^{\sigma}\big)\nonumber\\
&+
u_{1\delta}u_{\tau}\textrm{Tr}\big(e^{2\alpha\gamma^5}
\gamma^{\delta}\gamma^{\mu}\gamma^{\tau}\gamma^{\nu}\gamma^{\xi}\gamma^{\rho}\gamma^{\eta}\gamma^{\sigma}\big)
+
u_{\delta}u_{3\tau}\textrm{Tr}\big(e^{-2\alpha\gamma^5}
\gamma^{\xi}\gamma^{\mu}\gamma^{\delta}\gamma^{\nu}\gamma^{\tau}\gamma^{\rho}\gamma^{\eta}\gamma^{\sigma}\big)
\nonumber\\
&+
u_{1\delta}u_{3\tau}\textrm{Tr}\big(
\gamma^{\delta}\gamma^{\mu}\gamma^{\xi}\gamma^{\nu}\gamma^{\tau}\gamma^{\rho}\gamma^{\eta}\gamma^{\sigma}\big)
+
u_{\delta}u_{34\tau}\textrm{Tr}\big(
\gamma^{\xi}\gamma^{\mu}\gamma^{\delta}\gamma^{\nu}\gamma^{\eta}\gamma^{\rho}\gamma^{\tau}\gamma^{\sigma}\big)
\Big],
\\
{\pazocal{N}}^{\mu\nu\rho\sigma}_{(m^2u^2)}&=m^2 \Big[ u_{1\delta}\Big(u_{\tau}\textrm{Tr}\big(e^{2\alpha\gamma^{5}}\gamma^{\delta}\gamma^{\mu}
\gamma^{\tau}\gamma^{\nu}
\gamma^{\rho}
 \gamma^{\sigma}
\big)+u_{3\tau}\textrm{Tr}\big(\gamma^{\delta}\gamma^{\mu}
\gamma^{\nu}\gamma^{\tau}
\gamma^{\rho}
 \gamma^{\sigma}
\big) \cr
&+u_{34\tau}\textrm{Tr}\big(e^{2\alpha\gamma^{5}}\gamma^{\delta}\gamma^{\mu}
\gamma^{\nu}
\gamma^{\rho}\gamma^{\tau}
 \gamma^{\sigma}
\big)\Big)+u_{\delta}u_{3\tau}\textrm{Tr}\big(e^{-2\alpha\gamma^{5}}\gamma^{\mu}\gamma^{\delta}
\gamma^{\nu}\gamma^{\tau}
\gamma^{\rho}
 \gamma^{\sigma}
\big)\cr
& +u_{\delta}u_{34\tau}\textrm{Tr}\big(\gamma^{\mu}\gamma^{\delta}
\gamma^{\nu}
\gamma^{\rho}\gamma^{\tau}
 \gamma^{\sigma}
\big)+u_{3\delta}u_{34\tau}\textrm{Tr}\big(e^{2\alpha\gamma^{5}}\gamma^{\mu}
\gamma^{\nu}\gamma^{\delta}
\gamma^{\rho}
\gamma^{\tau} \gamma^{\sigma}
\big)\Big],
\\
{\pazocal{N}}^{\mu\nu\rho\sigma}_{(u^4)}&= u_{1\delta}u_{\tau}u_{3\xi}u_{34\eta} \textrm{Tr}\big(e^{4\alpha\gamma^{5}}\gamma^{\delta}\gamma^{\mu}
\gamma^{\tau}\gamma^{\nu}
\gamma^{\xi}\gamma^{\rho}
\gamma^{\eta} \gamma^{\sigma}
\big),
\label{eq:app-5}
\\
{\pazocal{N}}^{\mu\nu\rho\sigma}_{(m^4)}&=m^4\textrm{Tr}\big(\gamma^{\mu}
\gamma^{\nu}
\gamma^{\rho}
 \gamma^{\sigma}
\big). \label{eq:ap6}
\end{align}


\global\long\def\link#1#2{\href{http://eudml.org/#1}{#2}}
 \global\long\def\doi#1#2{\href{http://dx.doi.org/#1}{#2}}
 \global\long\def\arXiv#1#2{\href{http://arxiv.org/abs/#1}{arXiv:#1 [#2]}}
 \global\long\def\arXivOld#1{\href{http://arxiv.org/abs/#1}{arXiv:#1}}

{}

\end{document}